# On the blackhole dynamic potentials

Koustubh Ajit Kabe

THE THEORY GROUP

*Department of Physics, Sir Sitaram and Lady Shantabai Patkar College of Arts & Science, S.V. Road, Goregaon (West), Mumbai – 400 062 INDIA.*

**Abstract.** In the following paper, certain blackhole dynamic potentials have been developed definitively on the lines of classical thermodynamics. These potentials have been refined in view of the small differences in the equations of the laws of *blackhole dynamics* as given by Bekenstein and those of thermodynamics. Nine fundamental blackhole dynamical relations have been developed akin to the four fundamental thermodynamic relations of Maxwell. The specific heats $C_{\Omega,\Phi}$ and $C_{J,Q}$ have been defined. For a blackhole, these quantities are *negative*. The $\kappa\, dA$ equation has been obtained as an application of these fundamental relations. Time reversible processes observing constancy of surface gravity are considered and an equation connecting the internal energy of the blackhole $E$, the additional available energy defined as the first free energy function $K$, and the surface gravity $\kappa$, has been obtained. Finally as a further application of the fundamental relations, it has been proved for a *homogeneous gravitational field* in *blackhole spacetimes* that

$$C_{\Omega,\Phi} - C_{J,Q} = \kappa \left[ \left(\frac{\partial J}{\partial \kappa}\right)_{\Omega,\Phi} \left(\frac{\partial \Omega}{\partial \kappa}\right)_{J,Q} + \left(\frac{\partial Q}{\partial \kappa}\right)_{\Omega,\Phi} \left(\frac{\partial \Phi}{\partial \kappa}\right)_{J,Q} \right]$$

. This is dubbed as the *homogeneous fluid approximation* in context of the blackholes.

The classical theory of relativistic gravity ─ the general theory of relativity ─ asserts that a gravitationally collapsing star of mass $M$ will shrink, in short time as measured by an observer on the surface, to a radius of the order of $2GM/c^2$, known as its *gravitational radius* or *Schwarzschild radius*, at which the gravitational field becomes so strong that no further radiation or anything else can escape to infinity, the region of spacetime from which it is not possible to escape to infinity is a blackhole. The boundary of the blackhole, called as the *event horizon*, is an outgoing null *hypersurface* that just fails to reach infinity. The existence of finite entropy [1] as predicted for the blackholes implied the possibility of radiating blackholes [2], in turn proven conclusively by the quantum considerations of a blackhole neighborhood. We now have a standard result verified many times (see for example [3] and references therein) that blackholes emit radiation, now known as the Hawking Radiation. The noteworthy feature of the Hawking radiation tunneling through the event horizon of the blackhole is that it possesses an exact thermal spectrum. Consequently, it is observed that the dynamics of blackholes runs competitively on the same lines as that of classical thermodynamics. The thermal radiation emitted by a blackhole corresponds to a temperature of

$$T_H = \frac{\kappa h}{2\pi k_B c}, \tag{1}$$

where $\kappa$ is the surface gravity of the blackhole given by

$$\kappa = \frac{4\pi (r_+ c^2 - GM)}{A}. \tag{2}$$

This once again double implies the premise of finite entropy which expressed by the Bekenstein-Hawking formula:

$$S_{BH} = \frac{k_B}{4 l_{Pl}^2} A, \tag{3}$$

where $l_{Pl}$ is the Planck's length expressed by $l_{Pl} = \sqrt{\frac{Gh}{c^3}} \approx 10^{-33} cm$.

In the above and henceforth $A$ is the area of the event horizon of the blackhole mathematically defined as

$$A = \frac{4\pi G}{c^4} \left( 2GM^2 - Q^2 + 2\sqrt{G^2 M^4 - J^2 c^2 - GM^2 Q^2} \right). \tag{4}$$

The quantity $r_+$ in eq(2) is defined by the relation:

$$r_+ = \frac{1}{c^2}\left( GM + \sqrt{G^2 M^2 - \frac{J^2 c^2}{M^2} - GQ^2} \right). \tag{5}$$

The solution of the coupled Einstein-Maxwell field equations

$$R_{ab} - \frac{1}{2} g_{ab} R = -\frac{8\pi G}{c^4} T_{ab}, \tag{6}$$

$$F^{ab}_{;b} = 4\pi J^a, \tag{7}$$

and

$$F_{ab;c} + F_{bc;a} + F_{ca;b} = 0, \tag{8}$$

subject to the constraints imprinted by *mass $M$, angular momentum $J$, and charge $Q$*, (at radial infinity) and subject to the existence of a physically non-singular horizon describes the *Kerr-Newman geometry* of a blackhole with mass $M$, angular momentum $J$ and charge $Q$. For a general Kerr-Newman blackhole, $\Omega$, the angular frequency of rotation of the hole is given by

$$\Omega = \frac{4\pi J}{MA}, \tag{9}$$

and $\Phi$, the potential of the event horizon is given by

$$\Phi = \frac{4\pi Q r_+}{A}. \tag{10}$$

The theoretical foundations of the subject of blackhole dynamics were laid by the works of Bekenstein [1], Hawking [2], Bardeen, Carter and others [4]; systematic and complete treatment is given by Straumann [5]. The statistical theory of internal (unobservable) configurations and micro-canonical ensemble in the blackhole framework has been developed and discussed in considerable detail by Hawking [6]. As a consequence to the theory of blackhole dynamics, the possibility of developing *blackhole dynamic potentials*, mutually associating them and developing fundamental relations between the blackhole parameters: $M, J, Q, \kappa, \Omega \, and \, \Phi$, with the use of these potentials becomes inevitable and feasible.

Any Kerr-Newman blackhole i.e., a blackhole of given mass, angular momentum and charge can have a large number $\sigma$ of different unobservable internal configurations which reflect the possible different initial configurations of the matter which collapsed to produce the hole. The logarithm of this number $\sigma$ can be regarded as the entropy of the blackhole and is a measure of the amount of information about the initial state, which was lost in the formation of the blackhole. This is written mathematically as

$$S_H = \ln \sigma .  \tag{11}$$

Bekenstein suggested that the area of the event horizon of the blackhole is a measure of its entropy which is what is embodied in eq(3). Further, Bekenstein was the first to suggest that some multiple of $\kappa$ should be regarded as representing in some sense, the temperature of the blackhole. He noted that energy is conserved for blackholes as well as it is for other phenomena that occur in the universe. This is personified in the first law of blackhole dynamics,

$$dE = d(M c^2) = \frac{\kappa c^2}{8 \pi G} dA + \Omega \, dJ + \Phi \, dQ \tag{12}$$

which connects the difference in the energy of two nearby blackhole equilibrium states to the differences in the area $A$ of the event horizons, in the angular momentum $J$, and in the charge $Q$. This is very similar to the first law of thermodynamics

$$dE = Tds - p \, dv \tag{13}$$

suggesting as already mentioned that one should regard some multiple of $A$ as the entropy of a blackhole.

Given the background as thus, we now proceed to the theoretical considerations of the paper. Consider eq(12); we have

$$\left( \frac{\partial E}{\partial A} \right)_{J,Q} = \frac{\kappa c^2}{8 \pi G}, \tag{14}$$

$$\left( \frac{\partial E}{\partial J} \right)_{A,Q} = \Omega, \tag{15}$$

and

$$\left( \frac{\partial E}{\partial Q} \right)_{A,J} = \Phi. \tag{16}$$

Since $dE$ is a perfect differential,

$$\left( \frac{\partial}{\partial J} \left( \frac{\partial E}{\partial A} \right)_{J,Q} \right)_{A,Q} = \left( \frac{\partial}{\partial A} \left( \frac{\partial E}{\partial J} \right)_{A,Q} \right)_{J,Q}, \tag{17}$$

using eq(14) and (15) we have
$$\left(\frac{\partial \kappa}{\partial J}\right)_{A,Q} = \frac{8\pi G}{c^2}\left(\frac{\partial \Omega}{\partial A}\right)_{J,Q}. \tag{18}$$

Next, we have
$$\left(\frac{\partial}{\partial Q}\left(\frac{\partial E}{\partial A}\right)_{J,Q}\right)_{A,J} = \left(\frac{\partial}{\partial A}\left(\frac{\partial E}{\partial Q}\right)_{A,J}\right)_{J,Q}, \tag{19}$$

using eq(14) and (16) we get
$$\left(\frac{\partial \kappa}{\partial Q}\right)_{A,J} = \frac{8\pi G}{c^2}\left(\frac{\partial \Phi}{\partial A}\right)_{J,Q}, \tag{20}$$

and finally,
$$\left(\frac{\partial}{\partial Q}\left(\frac{\partial E}{\partial J}\right)_{A,Q}\right)_{A,J} = \left(\frac{\partial}{\partial J}\left(\frac{\partial E}{\partial Q}\right)_{A,J}\right)_{A,Q}, \tag{21}$$

using eq(15) and (16) we have
$$\left(\frac{\partial \Omega}{\partial Q}\right)_{A,J} = \left(\frac{\partial \Phi}{\partial J}\right)_{A,Q}. \tag{22}$$

For any process in which the surface gravity $\kappa$ is constant and by zeroth law of blackhole dynamics (which states in essence that *for a stationary axisymmetric blackhole in a spacetime which is asymptotically flat, it is possible to give a general definition of the surface gravity $\kappa$ such that $\kappa$ is constant on the horizon*), for a stationary axisymmetric blackhole, we once again consider eq(12),

$$\frac{\kappa c^2}{8\pi G} dA = d\left(\frac{\kappa c^2}{8\pi G} A\right), \tag{23}$$

therefore,
$$d\left[E - \frac{\kappa c^2}{8\pi G} A\right] = \Omega\, dJ + \Phi\, dQ, \tag{24}$$

or
$$dK = \Omega\, dJ + \Phi\, dQ. \tag{25}$$

$K = E - \dfrac{\kappa c^2}{8\pi G} A$ is the additional available energy. Now blackholes have negative specific heat. Thus $K$ must observe the condition $K < \dfrac{1}{4} E$ in order that the blackholes be in a state of stable thermal equilibrium.

Now from eq(25),
$$\left(\frac{\partial K}{\partial J}\right)_Q = \Omega, \tag{26}$$

$$\left(\frac{\partial K}{\partial Q}\right)_J = \Phi .\tag{27}$$

since $dK$ is a perfect differential,

$$\left(\frac{\partial}{\partial Q}\left(\frac{\partial K}{\partial J}\right)_Q\right)_J = \left(\frac{\partial}{\partial J}\left(\frac{\partial K}{\partial Q}\right)_J\right)_Q ,\tag{28}$$

or using eq(26) and (27) we have

$$\left(\frac{\partial \Omega}{\partial Q}\right)_J = \left(\frac{\partial \Phi}{\partial J}\right)_Q .\tag{29}$$

Eq(29) is a more flexible version of eq(22), since the constancy of $A$ is not required and yet the equation still holds good.

We still have three relations: eq(18), (20) and (22). Now since in devouring matter and emitting thermal radiation, heat is involved in that there is net absorption of heat in the former case and rejection of heat in the latter case, we conclude that a blackhole has enthalpy corresponding to a Hawking temperature $T_H$ given by eq(1). We define the blackhole enthalpy mathematically as

$$H = E - \Omega J - \Phi Q ,\tag{30}$$

therefore,

$$dH = dE - \Omega dJ - \Phi dQ - J d\Omega - Q d\Phi ,\tag{31}$$

or using eq(12) we have

$$dH = \frac{\kappa c^2}{8\pi G} dA - J d\Omega - Q d\Phi .\tag{32}$$

Hence,

$$\left(\frac{\partial H}{\partial A}\right)_{\Omega,\Phi} = \frac{\kappa c^2}{8\pi G} ,\tag{33}$$

$$\left(\frac{\partial H}{\partial \Omega}\right)_{A,\Phi} = -J ,\tag{34}$$

$$\left(\frac{\partial H}{\partial \Phi}\right)_{A,\Omega} = -Q .\tag{35}$$

Since $dH$ is a perfect differential,

$$\left(\frac{\partial}{\partial \Omega}\left(\frac{\partial H}{\partial A}\right)_{\Omega,\Phi}\right)_{A,\Phi} = \left(\frac{\partial}{\partial A}\left(\frac{\partial H}{\partial \Omega}\right)_{A,\Phi}\right)_{\Omega,\Phi} ,\tag{36}$$

using eq(33) and (34) we have,

$$\left(\frac{\partial \kappa}{\partial \Omega}\right)_{A,\Phi} = -\frac{8\pi G}{c^2}\left(\frac{\partial J}{\partial A}\right)_{\Omega,\Phi} .\tag{37}$$

Again, considering the next pair,

$$\left(\frac{\partial}{\partial \Phi}\left(\frac{\partial H}{\partial A}\right)_{\Omega,\Phi}\right)_{A,\Omega} = \left(\frac{\partial}{\partial A}\left(\frac{\partial H}{\partial \Phi}\right)_{A,\Omega}\right)_{\Omega,\Phi}, \tag{38}$$

using eq(33) and (35),

$$\left(\frac{\partial \kappa}{\partial \Phi}\right)_{A,\Omega} = -\frac{8\pi G}{c^2}\left(\frac{\partial Q}{\partial A}\right)_{\Omega,\Phi}. \tag{39}$$

And considering the final pair,

$$\left(\frac{\partial}{\partial \Omega}\left(\frac{\partial H}{\partial \Phi}\right)_{A,\Omega}\right)_{A,\Phi} = \left(\frac{\partial}{\partial \Phi}\left(\frac{\partial H}{\partial \Omega}\right)_{A,\Phi}\right)_{A,\Omega}, \tag{40}$$

using eq(34) and (35),

$$\left(\frac{\partial Q}{\partial \Omega}\right)_{A,\Phi} = \left(\frac{\partial J}{\partial \Phi}\right)_{A,\Omega}. \tag{41}$$

Eq(41) may not be considered as a fundamental relation as $Q$ under normal circumstances, does not relate to $\Omega$, and the same goes for $J$ and $\Phi$; yet we shall consider it as a fundamental relation in case in the future someone proves a relation between these quantities which was hitherto unknown due to the unobservable nature of the internal configurations of the blackhole and the quantum nature of singularity as well as that in the framework of supergravity [7].

Now consider eq(32). For constancy of surface gravity $\kappa$, angular frequency of rotation $\Omega$, and potential $\Phi$ of the event horizon, simultaneously,

$$d\left[H - \frac{\kappa c^2}{8\pi G}A\right] = 0. \tag{42}$$

We define an auxiliary free energy $F$ as

$$F = H - \frac{\kappa c^2}{8\pi G}A. \tag{43}$$

Therefore,

$$dF = 0. \tag{44}$$

Using eq(30) in eq(43) and differentiating we have

$$dF = dE - \Omega dJ - \Phi dQ - Jd\Omega - Qd\Phi - \frac{\kappa c^2}{8\pi G}dA - \frac{c^2 A}{8\pi G}d\kappa. \tag{45}$$

Substituting for $dE$ from eq(12) in eq(45) we have

$$dF = -\frac{c^2 A}{8\pi G}d\kappa - Jd\Omega - Qd\Phi. \tag{46}$$

Hence, we get

$$\left(\frac{\partial F}{\partial \kappa}\right)_{\Omega,\Phi} = -\frac{c^2 A}{8\pi G}, \tag{47}$$

$$\left(\frac{\partial F}{\partial \Omega}\right)_{\kappa,\Phi} = -J, \tag{48}$$

$$\left(\frac{\partial F}{\partial \Phi}\right)_{\kappa,\Omega} = -Q. \tag{49}$$

Since $dF$ is a perfect differential,

$$\left(\frac{\partial}{\partial \Omega}\left(\frac{\partial F}{\partial \kappa}\right)_{\Omega,\Phi}\right)_{\kappa,\Phi} = \left(\frac{\partial}{\partial \kappa}\left(\frac{\partial F}{\partial \Phi}\right)_{\kappa,\Phi}\right)_{\Omega,\Phi}, \tag{50}$$

using eq(48) and (49), we have

$$\left(\frac{\partial A}{\partial \Omega}\right)_{\kappa,\Phi} = \frac{8\pi G}{c^2}\left(\frac{\partial J}{\partial \kappa}\right)_{\Omega,\Phi}. \tag{51}$$

For the next pair, we have

$$\left(\frac{\partial}{\partial \Phi}\left(\frac{\partial F}{\partial \kappa}\right)_{\Omega,\Phi}\right)_{\kappa,\Omega} = \left(\frac{\partial}{\partial \kappa}\left(\frac{\partial F}{\partial \Phi}\right)_{\kappa,\Omega}\right)_{\Omega,\Phi}, \tag{52}$$

using eq(48) and (50)

$$\left(\frac{\partial A}{\partial \Phi}\right)_{\kappa,\Omega} = \frac{8\pi G}{c^2}\left(\frac{\partial Q}{\partial \kappa}\right)_{\Omega,\Phi}. \tag{53}$$

Finally, for the last pair we have

$$\left(\frac{\partial}{\partial \Omega}\left(\frac{\partial F}{\partial \Phi}\right)_{\kappa,\Omega}\right)_{\kappa,\Phi} = \left(\frac{\partial}{\partial \Phi}\left(\frac{\partial F}{\partial \Omega}\right)_{\kappa,\Phi}\right)_{\kappa,\Omega}, \tag{54}$$

using eq(49) and (50), we have

$$\left(\frac{\partial Q}{\partial \Omega}\right)_{\kappa,\Phi} = \left(\frac{\partial J}{\partial \Phi}\right)_{\kappa,\Omega}. \tag{55}$$

This completes our quest for the nine fundamental blackhole dynamic relations given by eqs (18),(20), (22) {or (29), (37), (39), (41), (51), (53) and (55). Finally, let us compare eq(14) and eq(33); we then have an auxiliary relation:

$$\left(\frac{\partial E}{\partial A}\right)_{J,Q} = \left(\frac{\partial H}{\partial A}\right)_{\Omega,\Phi}. \tag{56}$$

We now proceed to deduce an equation involving a $\frac{\kappa c^2}{8\pi G}dA$ term, dubbed as the $\kappa - dA$ equation in short.

We have

$$dA = \left(\frac{\partial A}{\partial \kappa}\right)_{\Omega,\Phi}d\kappa + \left(\frac{\partial A}{\partial \Omega}\right)_{\kappa,\Phi}d\Omega + \left(\frac{\partial A}{\partial \Phi}\right)_{\kappa,\Omega}d\Phi, \tag{57}$$

$$\frac{\kappa c^2}{8\pi G}dA = \frac{\kappa c^2}{8\pi G}\left(\frac{\partial A}{\partial \kappa}\right)_{\Omega,\Phi}d\kappa + \frac{\kappa c^2}{8\pi G}\left(\frac{\partial A}{\partial \Omega}\right)_{\kappa,\Phi}d\Omega + \frac{\kappa c^2}{8\pi G}\left(\frac{\partial A}{\partial \Phi}\right)_{\kappa,\Omega}d\Phi. \tag{58}$$

Now we define the specific heat of the blackhole at constant $\Omega, \Phi$ as

$$C_{\Omega,\Phi} = \frac{\kappa c^2}{8\pi G}\left(\frac{\partial A}{\partial \kappa}\right)_{\Omega,\Phi} \tag{59}$$

and using two of the above nine fundamental relations viz., eq(51) and (53) we have

$$\frac{\kappa c^2}{8\pi G} dA = C_{\Omega,\Phi}\, d\kappa + \kappa\left[\left(\frac{\partial J}{\partial \kappa}\right)_\Phi d\Omega + \left(\frac{\partial Q}{\partial \kappa}\right)_\Omega d\Phi\right]. \tag{60}$$

Similarly,

$$\frac{\kappa c^2}{8\pi G} dA = \frac{\kappa c^2}{8\pi G}\left(\frac{\partial A}{\partial \kappa}\right)_{J,Q} d\kappa + \frac{\kappa c^2}{8\pi G}\left(\frac{\partial A}{\partial \Omega}\right)_{J,Q} d\Omega + \frac{\kappa c^2}{8\pi G}\left(\frac{\partial A}{\partial \Phi}\right)_{J,Q} d\Phi. \tag{61}$$

we define the specific heat at constant $J$, $Q$ as

$$C_{J,Q} = \frac{\kappa c^2}{8\pi G}\left(\frac{\partial A}{\partial \kappa}\right)_{J,Q} \tag{62}$$

and using another pair of the nine fundamental relations likewise, we have

$$\frac{\kappa c^2}{8\pi G} dA = C_{J,Q}\, d\kappa + \kappa\left[\left(\frac{\partial J}{\partial \kappa}\right)_{A,Q} d\Omega + \left(\frac{\partial Q}{\partial \kappa}\right)_{A,J} d\Phi\right]. \tag{63}$$

As mentioned earlier, $C_{\Omega,\Phi} < 0$ and $C_{J,Q} < 0$.

The energy $E(=Mc^2)$ of the blackhole is in principle the sum of the additional available energy (or the first free energy function as we have called it) $K$ and $\frac{\kappa c^2}{8\pi G} dA$. The latter is what is called the bound energy or the energy that is not available for work. This makes $K$, that energy which is available for work in time-reversible processes (white holes) observing constancy of surface gravity. Since the area of the event horizon always tends to increase, it is clear that the bound energy of the blackhole always tends to increase, with the result that $K$, the additional available energy for work, tends to decrease. This decrease in $K$ tends the blackholes to approach a state of stable thermal equilibrium. Hence, the first free energy function $K$ is called the additional available energy.

Let us now consider what happens in the case of a time-reversible process observing constancy of surface gravity. For an infinitesimal time-reversible process,

$$dK = dE - d\left(\frac{\kappa c^2}{8\pi G} A\right) \tag{64}$$

or

$$dK = dE - \frac{\kappa c^2}{8\pi G} dA - \frac{c^2 A}{8\pi G} d\kappa. \tag{65}$$

Using eq(12), we have

$$dK = -\frac{c^2 A}{8\pi G} d\kappa + \Omega dJ + \Phi dQ \qquad (66)$$

Since the surface gravity is constant $d\kappa = 0$ and we have back our eq(25), viz.,
$$dK = \Omega dJ + \Phi dQ, \qquad (25)$$
or
$$K_2 - K_1 = \int_1^2 \Omega dJ + \int_1^2 \Phi dQ, \qquad (67)$$
i.e., the change in the additional available energy or the first free energy function of a blackhole, during a time-reversible process, observing constancy of surface gravity, is equal to the work done upon the blackhole system. In other words, the entire work in such a process is done at the cost of the additional available energy of the hole.

    Just as a blackhole performs work at the expense of its potential energy, so also, does a blackhole undergoing a time-reversible isothermal process performs work at the expense of its additional available energy. This of course is a hypothetical situation unless the blackhole is stationary (or at least quasi-stationary) and axisymmetric throughout any such time-reversible process that it undergoes.

    Given the entropy of a system as a function of the energy $E$ of the system and various other macroscopic parameters, one can define the temperature as $\frac{1}{T} = \frac{\partial S}{\partial E}$. Thus, one can define the temperature of the blackhole to be
$$\frac{1}{T_H} = \left(\frac{\partial S_H}{\partial E}\right)_{J,Q}. \qquad (68)$$
The generalized second law of blackhole dynamics as given by Bekenstein is then equivalent to the requirement that heat should not run uphill from a cooler system to a warmer one. With context to the radiation communicated by blackholes (Hawking radiation and other forms of quantum radiation decreasing the area of event horizon of the blackhole) this law is given in a more convenient statement [8]. We define the area of event horizon $A$ of a blackhole as the rate of change of the additional available energy with the surface gravity at constant angular momentum and charge, i.e,
$$A = -\frac{8\pi G}{c^2}\left(\frac{\partial K}{\partial \kappa}\right)_{J,Q} \qquad (69)$$
and we have from eq(24) and (25),
$$E = K + \frac{\kappa c^2}{8\pi G} A \qquad (70)$$
or
$$E = K - \left(\kappa \frac{\partial K}{\partial \kappa}\right)_{J,Q}. \qquad (71)$$
the main importance of the additional available blackhole energy lies in the field of statistical mechanics of blackholes.

The second free energy function $F$ is the blackhole dynamic potential at constant angular frequency of rotation and constant potential of the blackhole. We have seen that $F$ remains constant during a process in which the angular frequency of rotation $\Omega$ and potential $\Phi$ of the event horizon remain constant. It is also apparent that the blackhole enthalpy may be expressed as the sum of the second free energy function and the bound energy.

Lastly, we consider the homogeneous fluid approximation, which holds for homogeneous gravitational fields in blackhole spacetimes.

In this case $A$ is a function of surface gravity, angular momentum and charge. This is apparent from eq(12), or from eq(2) (9) and (10)

$$dA = \left(\frac{\partial A}{\partial \kappa}\right)_{J,Q} d\kappa + \left(\frac{\partial A}{\partial J}\right)_{\kappa,\Phi} d\Omega + \left(\frac{\partial A}{\partial Q}\right)_{\kappa,\Omega} d\Phi \quad , \tag{72}$$

therefore,

$$\left(\frac{\partial A}{\partial \kappa}\right)_{\Omega,\Phi} = \left(\frac{\partial A}{\partial \kappa}\right)_{J,Q} + \left(\frac{\partial \kappa}{\partial \kappa}\right)_{J,Q} + \left(\frac{\partial A}{\partial \Omega}\right)_{\kappa,\Phi}\left(\frac{\partial \Omega}{\partial \kappa}\right)_{J,Q} + \left(\frac{\partial A}{\partial \Phi}\right)_{\kappa,\Omega}\left(\frac{\partial \Phi}{\partial \kappa}\right)_{J,Q}. \tag{73}$$

Using two of the fundamental relations viz., eq(51) and (53), we conceive

$$\frac{\kappa c^2}{8\pi G}\left(\frac{\partial A}{\partial \kappa}\right)_{\Omega,\Phi} = \frac{\kappa c^2}{8\pi G}\left(\frac{\partial A}{\partial \kappa}\right)_{J,Q} + \kappa\left[\left(\frac{\partial J}{\partial \kappa}\right)_{\Omega,\Phi}\left(\frac{\partial \Omega}{\partial \kappa}\right)_{J,Q} + \left(\frac{\partial Q}{\partial \kappa}\right)_{\Omega,\Phi}\left(\frac{\partial \Phi}{\partial \kappa}\right)_{J,Q}\right], \tag{74}$$

or using eq(59) and (62) we have

$$C_{\Omega,\Phi} - C_{J,Q} = \kappa\left[\left(\frac{\partial J}{\partial \kappa}\right)_{\Omega,\Phi}\left(\frac{\partial \Omega}{\partial \kappa}\right)_{J,Q} + \left(\frac{\partial Q}{\partial \kappa}\right)_{\Omega,\Phi}\left(\frac{\partial \Phi}{\partial \kappa}\right)_{J,Q}\right]. \tag{75}$$

This is the approximation that we needed. Wherever the difference of specific heats is needed, one needs to use eq(78) and induct the value given by the right hand side of eq(75). Since $C_{\Omega,\Phi}$ and $C_{J,Q}$ are both negative, the right hand side will be positive if and only if, $C_{J,Q}$ is more negative than $C_{\Omega,\Phi}$. On the other hand, if $C_{\Omega,\Phi}$ is relatively more negative, then the right hand side will be negative. For a Schwarzschild blackhole, both the specific heats are one and the same.

The nine fundamental relations deduced above might find applications in a wide range of problems in blackhole thermodynamics, especially in the physics of the accretion tori or accretion disks around blackholes and in the physics of radiating and primordial blackholes wherein temperatures are significant. These and many other applications in various areas are plausible.

───────────────────────────────

**References**
1. J.D. Bekenstein, Phys. Rev. **D 7**, 2333 (1973); *ibid.,* Phys. Rev. **D 9**, 3292 (1974).
2. S.W. Hawking, Nature. **248**, 30 (1974); *ibid.,* Commun. Math. Phys. **43,** 199 (1975) ); *ibid.,* Commun. Math. Phys. **25**, 152 (1972).
3. B.S. DeWitt, Phys. Rep. **19C**, 295 (1975).
4. J.M. Bardeen, B. Carter and S.W. Hawking, Commun. Math. Phys.**31**, 161(1973).
5. N. Straumann, General Relativity with applications to Astrophysics, (Springer, Berlin 2004).
6. S.W. Hawking, Phys. Rev. **D 13**, 191 (1976).
7. F.A. Bais, *Blackholes in Compactified Supergravity* in PROCEEDINGS OF THE JOHNS HOPKINS WORKSHOP ON CURRENT PROBLEMS IN PARTICLE THEORY 7 (Bonn, 1983).
8. K. A. Kabe, Phys. Rev. **D** (to be published).